\documentclass[twocolumn,showpacs,preprintnumbers,amsmath,amssymb,superscriptaddress]{revtex4}
\usepackage{amssymb}
\usepackage{graphicx,color}
\usepackage{bm}

\begin{document}

\title{Hadron production in p+p, p+Pb, and Pb+Pb collisions with the HIJING 2.0 model at energies available at the CERN Large Hadron Collider }

\author{Wei-Tian Deng}
\affiliation{Frankfurt Institute for Advanced Studies (FIAS)
Ruth-Moufang-Strasse 1, D-60438 Frankfurt am Main, Germany}
\affiliation{Department of Physics, Shandong University, Jinan 250100, China}

\author{Xin-Nian Wang}
\affiliation{Institute of Particle Physics, Huazhong Normal University, Wuhan 430079, China}
\affiliation{Nuclear Science Division, MS 70R0319,
Lawrence Berkeley National Laboratory, Berkeley, California 94720}

\author{Rong Xu}
\affiliation{Institute of Particle Physics, Huazhong Normal University, Wuhan 430079, China}

\begin{abstract}
The HIJING (Heavy-ion Jet Interaction Generator) Monte Carlo model is updated with the latest parton distributions functions (PDF)
and new set of the parameters in the two-component mini-jet model that controls total $p+p$ cross section and the central
pseudorapity density. We study hadron spectra and multiplicity distributions using the HIJING 2.0 model and compare to recent
experimental data from $p+p$ collisions at the LHC energies. We also give predictions of hadron production in $p+p$, $p+Pb$ and $Pb+Pb$ collisions at the full LHC energy.
\end{abstract}


\maketitle

\section{Introduction}

The HIJING (Heavy Ion Jet INteraction Generator) Monte Carlo model was developed to study hadron production in high-energy
nucleon-nucleon, nucleon-nucleus and nucleus-nucleus collisions \cite{Wang:1991hta,Gyulassy:1994ew,Wang:1996yf}. It was based on a two-component
geometrical model of mini-jet production and soft interaction and has incorporated nuclear effects such as nuclear modification
of the parton distribution functions and jet quenching via final state jet medium interaction.  It can reproduce most features of
hadron production in $p+p(\bar p)$ and $p+A$ collisions up to the Fermilab Tevatron energies. With some modification of the string
configuration in the soft sector of particle production, it can also reproduce the bulk hadron spectra  and the approximately the suppression of high $p_{T}$ hadrons due to jet quenching in  central rapidity region of $A+A$ collisions up to the RHIC
energies \cite{Vance:1998vh,Vance:1999pr,Pop:2004dq}. It has been widely used to simulate hadron production in $p+A$ and $A+A$ collisions for designs of new detector systems and provide initial conditions for parton and hadron cascade models
such AMPT model \cite{Zhang:1999bd}.

The core component of HIJING is the two-component model for beam parton interaction which was proposed to
model the energy dependence of total cross section\cite{Gaisser:1984pg, L'Heureux:1985jk,Pancheri:1986qg,Durand:1987yv,DiasdeDeus:1987yw,Hwa:1987mm,Capella:1979fm} and particle
production\cite{Gaisser:1985jb,Sjostrand:1987su,Chen:1988bx, Wang:1990qp}
in high energy hadron collisions. The two-component model was also extended \cite{Blaizot:1987nc, Kajantie:1987pd} and later
incorporated in the HIJING model to describe initial parton production in high-energy heavy-ion collisions.
In this two-component model,
one assumes that parton interaction in high-energy nucleon-nucleon collisions can be divided
into soft interaction and hard or semi-hard
interaction with jet production. A cut-off scale $p_0$ in the transverse momentum of the final jet
production has to be introduced below
which the interaction is considered non-perturbative and can only be characterized by a finite
soft parton cross section $\sigma_{soft}$.
For jet production with transverse momentum $p_T>p_0$, the cross section and jet spectrum
are assumed to be given by perturbative
QCD parton model. Jet cross sections in collinear factorized perturbative QCD in
turn depend on the parton distribution functions (PDF)
that are parameterized from a global fit to the available experimental data of deeply
inelastic scattering (DIS) of lepton and nucleon,
Drell-Yan lepton pair, direct photon and jet production in $p+p(\bar p)$ collisions.

The two parameters $\sigma_{soft}$ and $p_0$ in HIJING are
determined phenomenologically by fitting the experimental data of total cross sections and hadron multiplicity in $p+p/\bar{p}$
collisions, which should depend on the parameterization of nucleon PDF's. In the original version of HIJING,
Duke-Own parameterization \cite{Duke:1983gd}  of PDF's are used to calculate the jet production cross
section with $p_{T}>p_{0}$ which are adequate for
description of jet production at RHIC and maybe Tevatron energies. However, for $p+p$ collisions at much higher energies such
as the LHC energies, jet production processes involve initial beam partons with fraction momentum  $x\sim 10^{-4}$ where
Duke-Owen parameterization of PDF's is no longer valid. More modern parameterizations should be used. In this
progress report,  we introduce the HIJING 2.0 model in which we use more modern set of parameterized PDF's. We have to
perform a global fit of the total cross sections and hadron multiplicities in $p+p (\bar p)$ collisions to determine the
two parameters, $\sigma_{soft} $ and $p_{0}$, and their energy dependence. We will use the new HIJING 2.0 model to
study hadron production in $p+p$ at the LHC energies and compare to the recently published experimental data. We also provide
predictions of hadron multiplicities in $p+p$ and $Pb+Pb$ collisions at the full LHC energies.

\section{Two-component model in HIJING}

In the two-component model\cite{Wang:1991hta,Gyulassy:1994ew,Wang:1996yf,Gaisser:1984pg, L'Heureux:1985jk,Pancheri:1986qg, Durand:1987yv,DiasdeDeus:1987yw,Hwa:1987mm,Capella:1979fm}
incorporated in HIJING, one assumes that events of nucleon-nucleon collisions at high energy can be divided into soft and
hard processes with at least on pair of jet production with $p_{T}>p_{0}$. The inclusive jet
cross section $\sigma_{jet}$ in the leading order (LO) \cite{Eichten:1984eu}
\begin{equation}
\label{eq:sigma_jet}
 \sigma_{jet}=\int_{p_0^2}^{s/4}\mathrm{d}p_T^2\mathrm{d}y_1\mathrm{d}y_2
 \frac{1}{2}\frac{\mathrm{d}\sigma_{jet}}{\mathrm{d}p_T^2 \mathrm{d}y_1 \mathrm{d}y_2},
\end{equation}
\begin{equation}
 \frac{\mathrm{d}\sigma_{jet}} {\mathrm{d}p_T^2 \mathrm{d}y_1 \mathrm{d}y_2} =K \sum_{a,b}x_1f_a(x_1,p_T^2)x_2f_b(x_2,p_T^2) \frac{\mathrm{d}\sigma^{ab}(\hat{s},\hat{t},\hat{u})}{\mathrm{d}\hat{t}}
\end{equation}
depends on the parton-parton cross section $\sigma^{ab}$ and parton distribution functions $f_a(x,p_T^2)$
in a parameterized form of experimental data of other high-energy DIS and nucleon-nucleon collisions, where the summation runs over all
parton species, $y_1$ and $y_2$ are the rapidities of the scattered partons, $x_1$ and $x_2$ are the light-cone momentum
fractions carried by the initial partons, and $K\approx2$ accounts for the next-to-leading order (NLO)
corrections to the leading order
jet cross section.

Within the eikonal formalism\cite{Gaisser:1984pg, L'Heureux:1985jk, Pancheri:1986qg, Durand:1987yv, DiasdeDeus:1987yw, Hwa:1987mm,Capella:1983ct}, the nucleon-nucleon cross sections can be expressed in the impact-parameter representation as
\begin{eqnarray}
\label{eq:eikonal_el}
 \sigma_{el}&=&\pi\int_0^{\infty}\mathrm{d}b^2\left[1-e^{\chi(b,s)}\right] ^2,\\
\label{eq:eikonal_in}
 \sigma_{in}&=&\pi\int_0^{\infty}\mathrm{d}b^2\left[1-e^{2\chi(b,s)}\right] ,\\
\label{eq:eikonal_tot}
 \sigma_{tot}&=&2\pi\int_0^{\infty}\mathrm{d}b^2\left[1-e^{\chi(b,s)}\right] ,
\end{eqnarray}
in the limit that the real part of the parton-parton scattering amplitude can be neglected. The eikonal functions $\chi(b,s)$
at an impact parameter $b$ is therefore real and can be expressed in term of inclusive jet cross section in pQCD and an
effective cross section for the non-perturbative soft parton-parton collisions within the two-component model  \cite{Wang:1991hta},
\begin{eqnarray}
\nonumber
 \chi(b,s)&\equiv& \chi_s(b,s)+\chi_h(b,s)\\
&&= \frac{1}{2} \left[\sigma_{soft}T_{NN}(b)+\sigma_{jet}T_{NN}(b)\right],
\end{eqnarray}
where $T_{NN}(b)$ is the nucleon-nucleon overlap function and will be assumed to take the form of the Fourier transformation
of a dipole form factor in HIJING.

Within the above eikonal implementation of the two-component model, one can also calculate the nucleon-nucleon cross
section \cite{Wang:1990qp} for no jet and $j\geq1$ number of jet production with $p_T>p_0$,
\begin{eqnarray}
\sigma_0&=& \pi\int_0^\infty \mathrm{d}b^2 \left[1-e^{-2\chi_s(b,s)}\right]e^{-2\chi_h(b,s)},\\
 \sigma_j&=& \pi\int_0^\infty \mathrm{d}b^2 \frac{\left[2\chi_h(b,s)\right]^j}{j!}e^{-2\chi_h(b,s)}.
 \label{eq:problt_jet}
\end{eqnarray}
Their sum gives rise to the total inelastic cross section in Eq.(\ref{eq:eikonal_in}). The above eikonal formalism is the base
for the Monte Carlo simulation of multiple jet production in $p+p$, $p+A$ and $A+A$ collisions. Both $\sigma_{soft}$ and the
transverse momentum cut-off $p_0$ are considered parameters in the HIJING model and are fit to the total and inelastic
nucleon-nucleon cross sections and the hadron multiplicity density in the middle rapidity $y=0$ at each colliding energy.

One can find the detailed description of HIJING model for hadron production in $p+p$, $p+A$ and $A+A$
collisions in Refs. \cite{Wang:1991hta}.  The main features of HIJING are:

(1) Multiple mini-jet production are simulated according to the above eikonal formalism for each nucleon-nucleon collisions at
given impact parameter $b$. The kinematics of each pair of jets and the associated initial and final state radiation are simulated
using PYTHIA model\cite{Sjostrand:1987su}.

(2) Events without jet production (with $p_{T}>p_{0}$) and the underlying soft parton interaction in events with jet production are
modeled by excitation of quark-diquark strings with gluon kinks along the lines of the FRITIOF model \cite{Andersson:1986gw}
and the DPM
model\cite{Capella:1979fm, Ranft:1987xn}. In addition, multiple low-$p_T$ exchanges among the end point
constituents are included.

(4) A set of impact-parameter-dependent parton distribution functions is used to include nuclear modification of the parton
distribution functions inside nuclei.

(5) A simple model for jet quenching is used to study the effect of jet medium interaction in $A+A$ collisions \cite{Wang:1991xy}.


\section{Updates in HIJING}

In the default setting of HIJING 1.0 \cite{Wang:1991hta},
Duke-Owens parameterization\cite{Duke:1983gd} of PDFs in nucleons is used.
With Duke-Owens parameterization of PDFs, an energy independent cut-off scale $p_0=$2 GeV/$c$ and a constant soft
parton cross section $\sigma_{soft}=57\;mb$ are sufficient to reproduce the experimental data on total and inelastic cross sections
and the hadron central rapidity density in $p+p/\bar{p}$ collisions \cite{Wang:1991us}.
Duke-Owens parameterization of PDFs is known to be very
outdated and one needs to use more modern parameterizations from new global fit to experimental data, especially at the LHC
energies when mini-jet production reaches to very small-$x$ region of the parton distribution where gluon distribution is
much higher than Duke-Owens parameterization.

Furthermore, with a constant transverse momentum cut-off $p_{0}=2$ GeV/$c$ in HIJING 1.0, the total number of
min-jets per unit transverse area could exceed the limit
\begin{equation}
 \frac{T_{AA}(b)\sigma_{jet}}{\pi R_{A}^{2}}\le \frac{p_{0}^{2}}{\pi}
\end{equation}
for independent multiple jet production even in central $p+p$ collisions for sufficiently large inclusive
jet cross section at high colliding energies, where $T_{AA}(b)$ is the overlap function of $A+A$ collisions and
$\pi/p_{0}^{2}$ is the intrinsic transverse size of a mini-jet with transverse momentum $p_{0}$. Therefore,
one inevitably has to increase the value of the transverse momentum cut-off $p_{0}$ to ensure the applicability
of the underlying two-component model of independent multiple jet production in the HIJING model.

\begin{figure}
  \centering
 \includegraphics[width=0.45\textwidth]{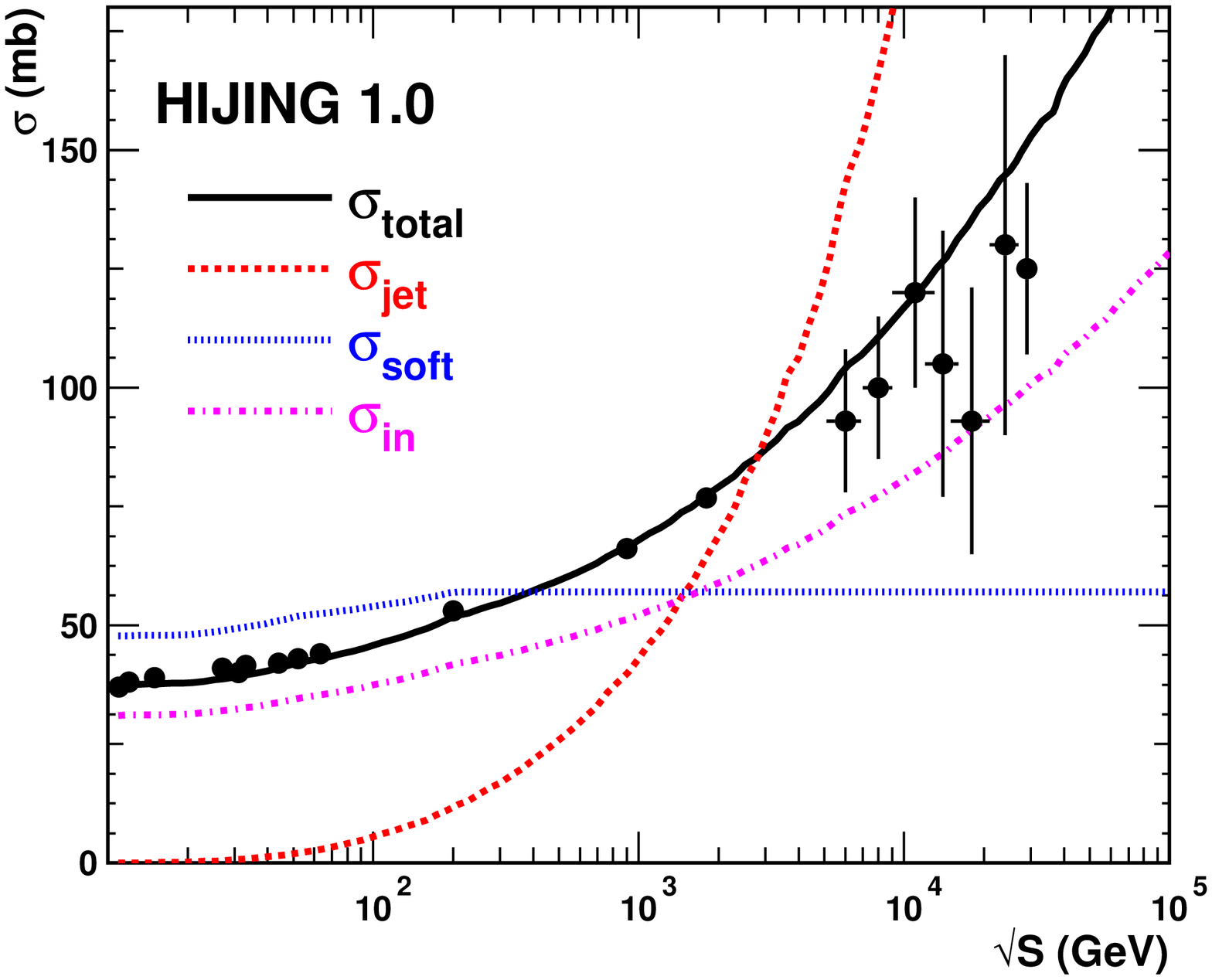}
 \includegraphics[width=0.45\textwidth]{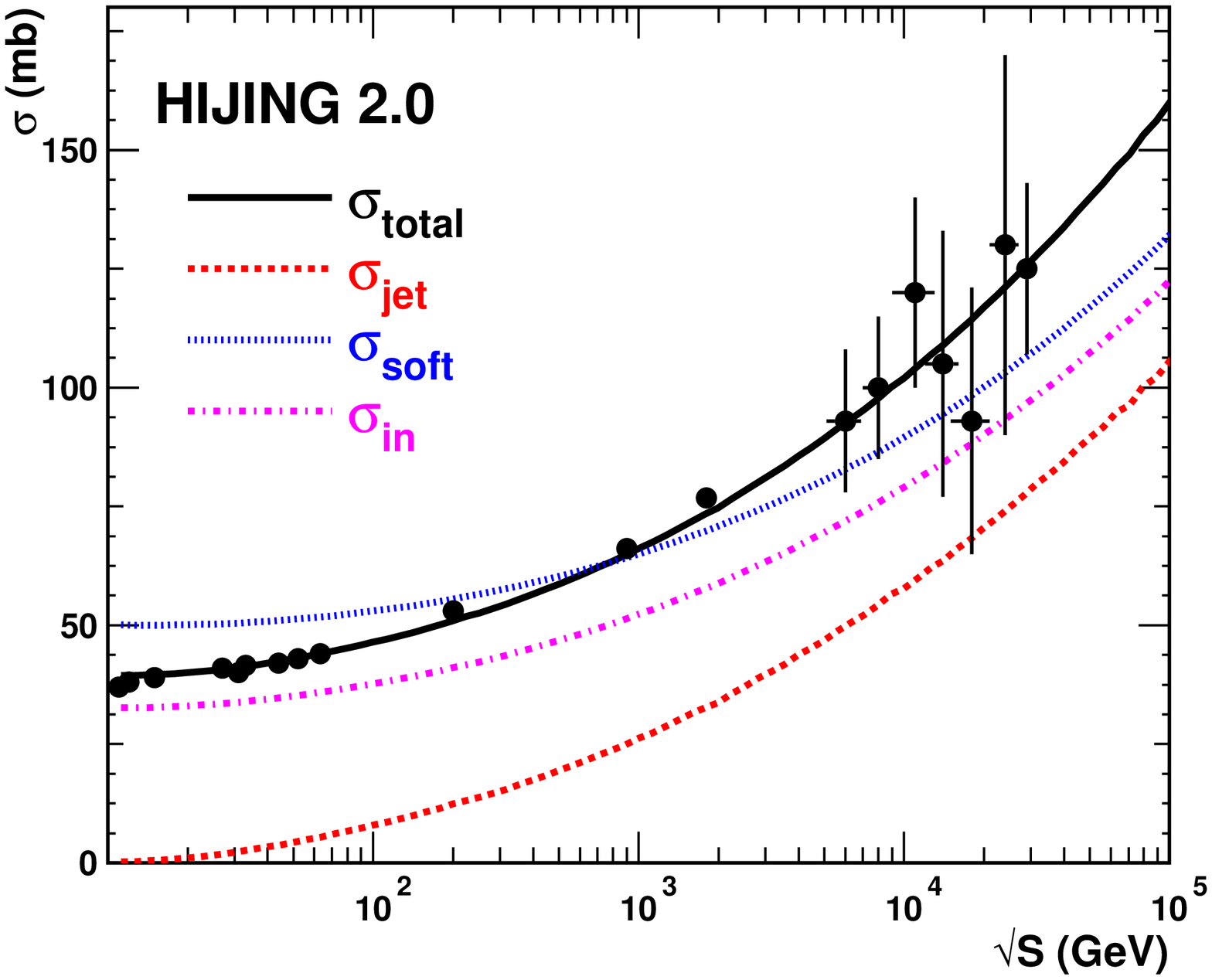}
  \caption{(color online) Total, soft and jet production cross sections of $pp$ and $p\bar{p}$ collisions. The
  histogram in left panel are calculated using HIJING 1.0, while the lower using HIJING2.0. The
  data are from \cite{Amaldi:1979kd, Bozzo:1984rk, Alner:1986iy, Amos:1989at, Baltrusaitis:1984ka, Hara:1983pa}.}
  \label{fig:cross_section_total}
\end{figure}

\begin{figure}
  \centering
 \includegraphics[width=0.4\textwidth]{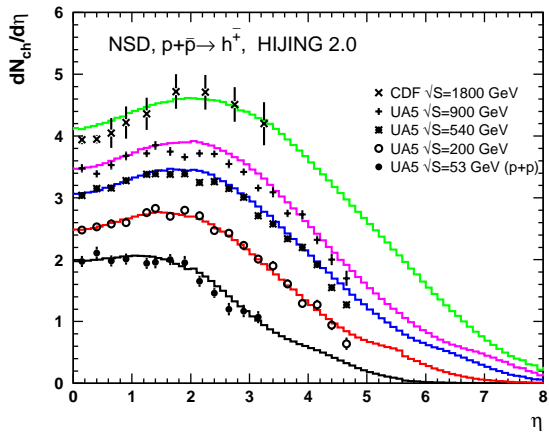}
  \caption{(color online) Pseudorapidity distributions of charged particles in non-single-diffractive $pp$ at $\sqrt{s}=53$ GeV, $p\bar{p}$ collisions at $\sqrt{s}=200,~540,~900,~1800$ GeV as compared to experimental data \cite{Alner:1986xu, Abe:1989td}.}
  \label{fig:dndeta_pp}
\end{figure}

In the updated version of HIJING 2.0, we will use the Gluck-Reya-Vogt(GRV) parameterization\cite{Gluck:1994uf} of PDFs,
among many modern parameterizations of the PDFs that are available. The gluon distributions in this new parameterization
is much higher than the old Duke-Owens parameterization at small $x$ and therefore give much larger inclusive jet cross
section at high colliding energies with a fixed value of cut-off $p_{0}$. One therefore can no longer fit the
experimental $p+p/\bar{p}$ data on total and inelastic cross sections using a constant cut-off $p_0$ and the soft
parton cross section $\sigma_{soft}$ within the two-component model. One has to assume an energy-dependent
cut-off $p_0(\sqrt{s})$ and soft cross section $\sigma_{soft}(\sqrt{s})$ \cite{Li:2001xa}. Fitting the experimental values of the total
and inelastic cross sections of $p+p/(\bar p)$ collisions including those extracted from cosmic experimental
and the hadron central rapidity density, we have the following parameterized energy-dependence of the cut-off and
the soft parton cross section used in the two-component model of HIJING 2.0:
\begin{eqnarray}
\label{eq:p0}
\nonumber
 p_0&=&2.62 -1.084\mathrm{log}(\sqrt{s})+0.299\mathrm{log}^2(\sqrt{s}) \\
&&-0.0292\mathrm{log}^3(\sqrt{s})+0.00151\mathrm{log}^4(\sqrt{s}),\\
\label{eq:sigma_soft}
\nonumber
 \sigma_{soft}&=& 55.316 -4.1126\mathrm{log}(\sqrt{s}) +0.854\mathrm{log}^2(\sqrt{s}) \\
&&-0.0307\mathrm{log}^3(\sqrt{s})+0.00328\mathrm{log}^4(\sqrt{s}),
\end{eqnarray}
where the colliding energy $\sqrt{s}$ in center-of-mass frame is in units of GeV.
Shown in Fig.~\ref{fig:cross_section_total} are the calculated total and inelastic cross sections using both HIJING 1.0 and
HIJING 2.0 as compared to the experimental data. The total inclusive jet cross section and non-perturbative soft parton
cross sections are also plotted for illustration. With a constant cut-off $p_{0}=$ 2 GeV/$c$ and soft parton cross
section $\sigma_{soft}$ at high colliding energies, HIJING 1.0 already gives larger total cross section than the cosmic
data indicate even with the Duke-Owens parameterization of PDFs. With much higher gluon distribution at small $x$
in the GRV parameterization used in HIJING 2.0, one has to introduce a cut-off $p_{0}$ and the soft parton cross
section $\sigma_{soft}$ that increase with colliding energy in order to fit the experimental data on the total cross
section. There are, however, some freedom in fixing the values of $p_{0}$ and $\sigma_{soft}$, which is further
constrained by the energy-dependence of the central rapidity density of the charged multiplicities, as shown in
Fig.~\ref{fig:dndeta_pp}. The increasing cut-off as required by the experimental data indicates that multiple
mini-jet production below such cut-off are no longer independent and coherent interaction becomes
important. This might be taken as an indirect evidence of gluon saturation at very small $x$ inside a proton
in proton-proton collisions at very high energies, especially at the LHC energies.  An alternative approach
to effectively take into account of such gluon saturation is to increase the string tension of soft hadron
production as proposed in Ref.~\cite{ToporPop:2007hb}. We choose to focus on the change of
minijet production in HIJING2.0.

\begin{figure}
  \centering
 \includegraphics[width=0.45\textwidth]{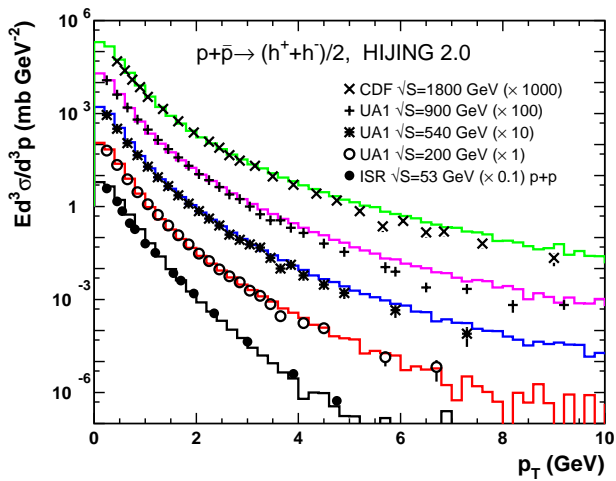}
  \caption{(color online) Invariant inclusive cross sections of charged particles in $pp$ at $\sqrt{s}=53$ GeV, $p\bar{p}$ collisions at $\sqrt{s}=200,~500,~900,~1800$ GeV. The data come from \cite{Alper:1974rw, Albajar:1989an, Abe:1988yu}. Both the calculation and experimental data are obtained in the region of $|\eta|<2.5$ for $\sqrt{s}=200,~540,~900$ GeV, and $|\eta|<1.0$ for $\sqrt{s}=53$ and $1800$ GeV.}
  \label{fig:dsigmadpt_pp}
\end{figure}

We also show in Fig.~\ref{fig:dsigmadpt_pp} the transverse momentum spectra
calculated with HIJING 2.0 at different colliding energies as compared to the experimental data. HIJING 2.0
results are all in good agreement with the experimental data.

Note that both HIJING 2.0  calculations and data shown in Fig.~\ref{fig:dndeta_pp} are for non-single-diffractive (NSD)
events. The definitions of NSD trigger are different in different experiments leading to different central pseudorapidity densities in
these NSD events.  The NSD triggers in UA5 experiment require at least one charged particle
simultaneously in each of the pseudorapidity regions at both ends covering $2<|\eta|<5.6$, while Collider-Detector
at Fermilab (CDF) NSD events are triggered in $3.2<|\eta|<5.9$. The increase of the central pseudorapidity density
with energy can be attributed to the increased mini-jet production in high colliding energies.



\section{Hadron spectra in $p+p$ and $A+A$ collisions at the LHC energies}

\begin{figure}
  \centering
 \includegraphics[width=0.45\textwidth]{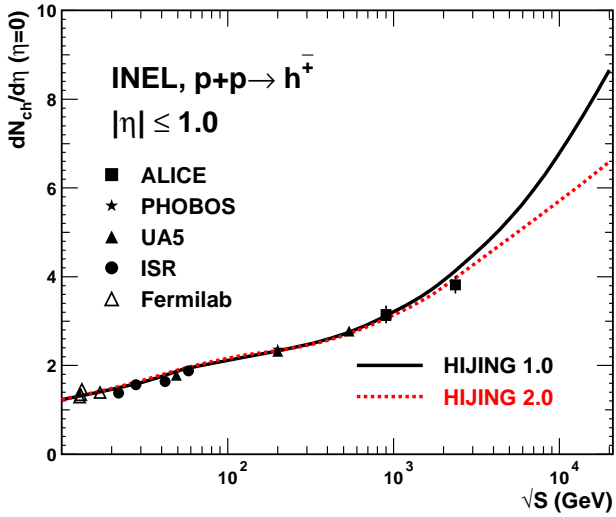}
  \caption{(color online) Central pseudorapidity density at $\eta =0$ for charged particles produced in $p+p/\bar{p}$ collisions
  as a function of $\sqrt{s}$. The experimental data are
  from \cite{Alner:1986xu,Thome:1977ky, Whitmore:1973ri, Nouicer:2004ke, :2009dt}.}
  \label{fig:dndet-s}
\end{figure}

With the updated HIJING 2.0, we can study hadron production in $p+p$ and $A+A$ collisions at the LHC energies. At the highest
energy of $\sqrt{s}=14$ TeV, mini-jet production involves partons in the very small $x$ region. The increase of the
cut-off for independent mini-jet production required to fit the total cross section can be considered as indication
of coherence or gluon saturation inside protons at very high energies. This will certainly affect the produced
hadron pseudorapidity distribution at the LHC energies. Shown in Fig.~\ref{fig:dndet-s} is the energy dependence
of the central pseudorapidity density of charged hadrons $dN_{\rm ch}/d\eta$ averaged over $|\eta|<1.0$ as a
function of the colliding energy in inelastic $p+p$ collisions from both HIJING1.0 and HIJING 2.0 calculations
as compared to experimental data, including that measured in $p+p$ collisions at $\sqrt{s}=0.9$ TeV and $2.36$
TeV from ALICE \cite{:2009dt} at the LHC.
The central rapidity hadron density in HIJING 2.0 continues to increase with the colliding energy at LHC
more or less in a double logarithm form. Because of the larger values of cut-off for mini-jet production, the central
rapidity hadron density in HIJING 2.0 is significantly lower than HIJING 1.0 at the LHC energies.

\begin{figure}
  \centering
 \includegraphics[width=0.45\textwidth]{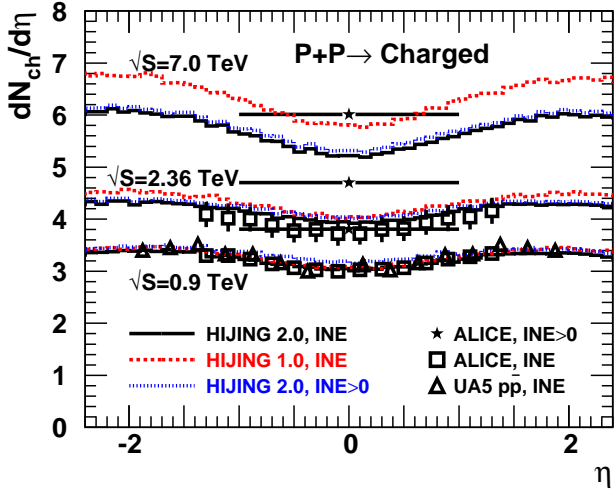}
  \caption{(color online) Pseudorapidity distribution $\mathrm{d}N_{ch}/\mathrm{d}\eta$ in inelastic $p+p$ collisions at LHC energies.
  The experiment data  are from UA5\cite{Alner:1986xu},  ALICE\cite{:2009dt} and  CMS\cite{Khachatryan:2010xs,Khachatryan:2010us}.
  Shown in short-dashed lines are the HIJING results for ALICE defined INEL$>$0 events in which at least one charged
  tracks are required within $|\eta|<1$. }
  \label{fig:dndeta_LHC1}
\end{figure}

\begin{figure}
  \centering
 \includegraphics[width=0.45\textwidth]{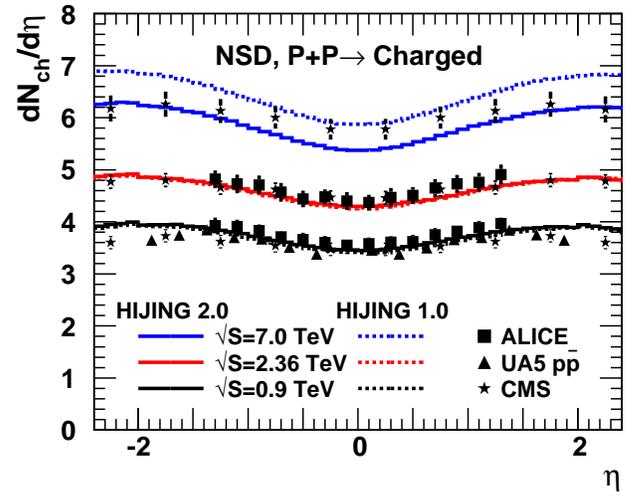}
  \caption{Pseudorapidity distribution $\mathrm{d}N_{ch}/\mathrm{d}\eta$ in non-single diffractive (NSD) $p+p$ collisions at LHC energies.
  The experiment data  are from UA5\cite{Alner:1986xu},  ALICE\cite{:2009dt} and  CMS\cite{Khachatryan:2010xs,Khachatryan:2010us}.
  The definitions of NSD events in our calculations and the experimental data from ALICE and UA5 follow that in
  UA5 experiment as described in the text. CMS \cite{Khachatryan:2010xs,Khachatryan:2010us} uses a different definition.}
  \label{fig:dndeta_LHC2}
\end{figure}

Shown in Fig.~\ref{fig:dndeta_LHC1} and \ref{fig:dndeta_LHC2} are HIJING results for the pseudorapidity distributions of charged hadrons
in $p+p$ at $\sqrt{s}=0.9$, 2.36 and 7 TeV in inelastic  and non-single diffractive (NSD) events. The central rapidity
density in NSD events is generally larger than in inelastic events and HIJING results agree with the experimental data
reasonably well. Note that CMS \cite{Khachatryan:2010xs,Khachatryan:2010us} experiment uses a different definition
of NSD events that gives slightly higher averaged multiplicity than that with UA5 definition of NSD events in
HIJING calculations. Another class of inelastic
events in ALICE experiments (INEL$>$0) is defined by requiring at least one charged tracks within $|\eta|<1$. HIJING
results for these type of events (short-dashed lines) are similar to the inelastic events because the lack of double-diffractive
events in the Mont Carlo model. Both single diffractive and non-diffractive events in HIJING have finite value of multiplicity
in the central rapidity region. This disagrees with the ALICE data \cite{:2009dt} which are even larger than the
central rapidity density in NSD events. Such discrepancy in trigger dependence of the averaged multiplicity can be
fixed when double-diffractive events are included in future improvement of HIJING. However, this is not expected
to have a significant influence on particle production in heavy-ion collisions because the fraction of diffractive interactions
becomes increasingly smaller in multiple nucleon-nucleon collisions.

\begin{figure}
  \centering
 \includegraphics[width=0.45\textwidth]{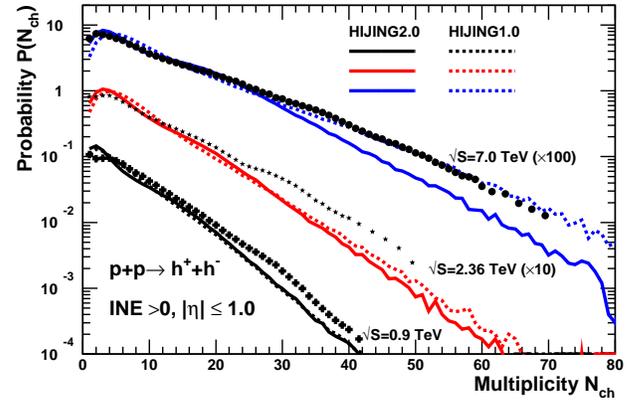}
  \caption{Multiplicity distributions for changed hadrons within $|\eta|<1$ in inelastic $p+p$ collisions at LHC energies from ALICE\cite{:2009dt} experiment  compared to HIJING results.}
    \label{fig:multi-distr}
\end{figure}

\begin{figure}
  \centering
 \includegraphics[width=0.45\textwidth]{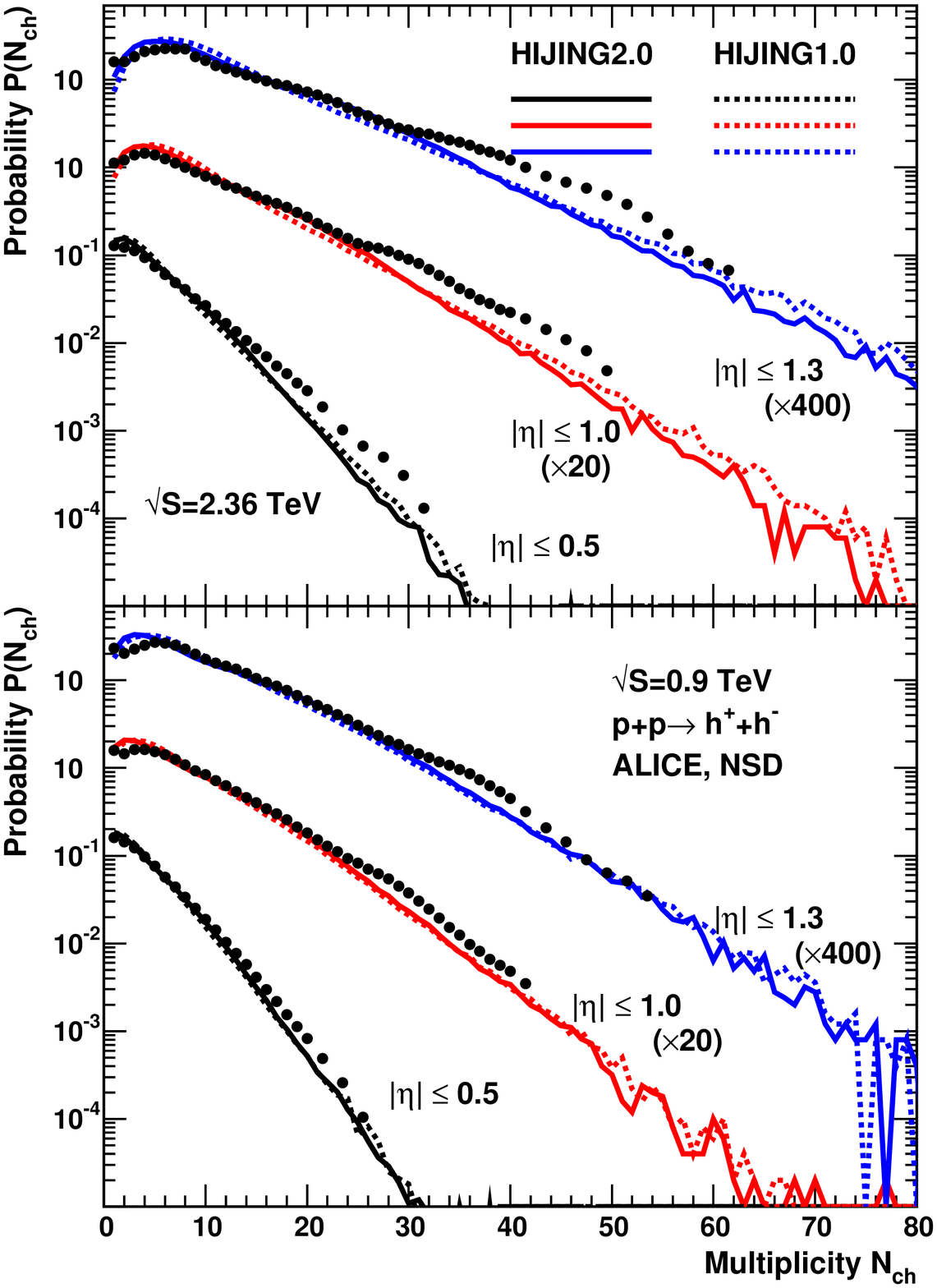}
  \caption{Multiplicity distributions for charged hadrons within different pseudorapidity intervals in NSD events of
  $p+p$ collisions at $\sqrt{s}=2.36$ TeV from ALICE\cite{:2009dt} experiment as compared to HIJING results.}
  \label{fig:multi-distr2}
\end{figure}

Shown in Figs.~\ref{fig:multi-distr} and \ref{fig:multi-distr2} are the multiplicity distributions for charged hadrons
within different pseudorapidity range in inelastic and NSD $p+p$ collisions at the LHC energies from HIJING
calculations as compared to experimental data from ALICE experiment \cite{:2009dt}. HIJING results agree
reasonably well with the data in low and intermediate multiplicities. However, they fall short of the experimental
data at higher multiplicity tails in particular at $\sqrt{s}=7$ TEV. These high multiplicity events are dominated by multiple jets and
they are likely to have final state interaction with each other as indicated by the ridge structure in the
two-hadron correlation in azimuthal angle and large rapidity gap observed in CMS
experiment \cite{Khachatryan:2010gv}. The final state interaction in these high multiplicity events
could affect the multiplicity distribution which is not included in current HIJING model.

\begin{figure}
  \centering
 \includegraphics[width=0.45\textwidth]{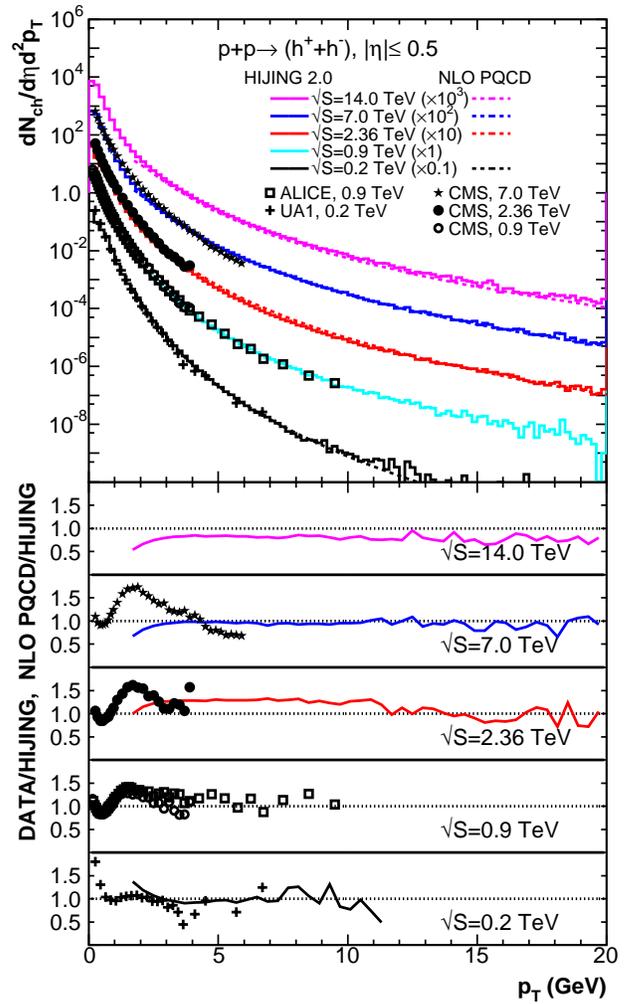}
  \caption{(color online) Transverse momentum distributions of charged hadron in $p+p$ collisions. The experimental
  data are from ALICE \cite{Aamodt:2010my} (open square) for inelastic $p+p$ collisions and CMS\cite{Khachatryan:2010xs}
  (open and closed circle) for NSD $p+p$ collisions at $\sqrt{s}=0.9$ and $2.36$ TeV. The histograms are
 results from HIJING 2.0.}
  \label{fig:dndpt_LHC}
\end{figure}

At high colliding energies, the transverse momentum of produced hadrons should be dominated by hadrons from
jet fragmentation, especially at large transverse momentum, as shown in  Fig.~\ref{fig:dndpt_LHC}. We compare
the HIJING 2.0 results to the ALICE \cite{Aamodt:2010my} experimental data on charged hadrons
within $|\eta|<0.8$ in inelastic $p+p$ collisions at $\sqrt{s}=0.9$ TeV and CMS \cite{Khachatryan:2010xs,Khachatryan:2010us}
data on charged hadrons within $|\eta|<2.4$ in NSD $p+p$ collisions at $\sqrt{s}=0.9$, $2.36$ and 7 TeV.
The HIJING 2.0 results at $\sqrt{s}=$ 14 TeV for inelastic events in $p+p$ collisions are also shown.
Within the LHC energy range  $\sqrt{s}=0.9 -14$ TeV, the transverse momentum spectra at large $p_{T}$
region becomes apparently harder at higher colliding energies.
Even though the cut-off for mini-jet production in HIJING 2.0 is significantly larger than in HIJING 1.0,
hadronization from partons as results of initial and final state radiation still fills in the hadron transverse momentum
spectra in the intermediate transverse momentum. HIJING 2.0 describes reasonably well the ALICE and CMS
experimental data at the LHC energies with the maximum discrepancy of 50\% at small $p_{T}\sim $ 2 GeV, as shown
in the lower panels in Fig.~\ref{fig:dndpt_LHC} where the ratios of data over HIJING results are plotted. The transverse
momentum spectra in this small $p_{T}$ region can also be influenced by final-state interaction among jets. We also
plot the ratios of charged hadron spectra from NLO pQCD calculation \cite{Harris:2001sx,Zhang:2007ja} over that
from HIJING. HIJING reproduces the NLO pQCD results very well over large $p_{T}$ range and at different
colliding energies.

\begin{figure}
  \centering
 \includegraphics[width=0.45\textwidth]{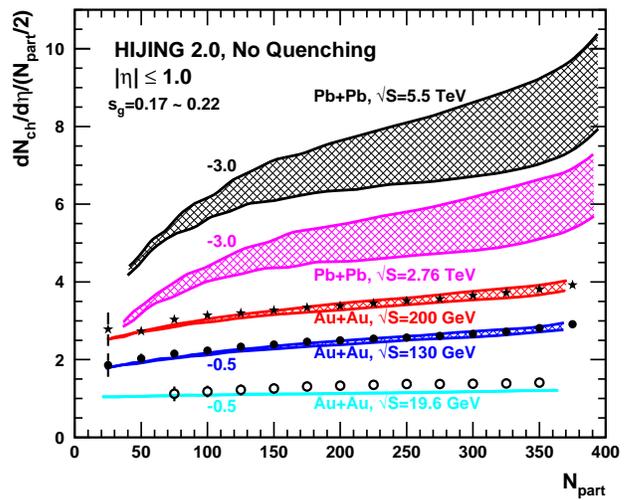}

  \caption{(color online) Pseudorapidity density per participant pair as a function of the average number of participants
  in heavy-ion collisions at different energies from HIJING2.0 calculations. Dashed lines are HIJING1.0 results.
  The experimental data are from Ref.~\cite{Adler:2004zn}. Note a vertical
  off-set is applied to each curve for clear illustration.}.
  \label{fig:dndeta-aa}
\end{figure}

To describe hadron production in $A+A$ collisions, one should include both the nuclear modification of the
parton distribution functions and jet quenching in final state interaction. Jet quenching in general suppresses
high transverse momentum hadrons \cite{Wang:1991xy}. Since the yields of  these high transverse momentum partons are
relatively small as compared to the bulk part of the initial parton production, its effect on the total hadron multiplicity in
$A+A$ collisions will be small as illustrated by other models
studies \cite{Wang:2000bf,Eskola:2000xq,Lin:2000cx,Kharzeev:2000ph} of hadron production in heavy-ion
collisions at RHIC, most of which have also neglected the effect of final state interaction. With two
parameters ($p_{0}$ and $\sigma_{soft}$) fixed from fit to $p+p$ collisions, the only uncertainty
for hadron multiplicity density in $A+A$ collisions is the nuclear modification of parton distribution
functions, gluon distributions in particular, at small $x$. We assume parton distributions in nuclei,
\begin{equation}
f_a^A(x,Q^2)=AR_a^A(x,Q^2)f_a^N(x,Q^2),
\end{equation}
are given by that in nucleon and the modification factor $R_a^A(x,Q^2)$.  This nuclear modification
has been studied with data from deeply inelastic scattering (DIS) and Drell-Yan lepton pair production
experiments. However, most of parameterizations \cite{Eskola:1998df, hkm} do not have gluon shadowing
strong enough to give the central multiplicity density as measured in RHIC experiments within the
two component model as implemented in HIJING model. We will use the HIJING new
parameterization \cite{Li:2001xa} which introduced a strong impact-parameter
dependence of the parton shadowing in order to fit the centrality dependence of the central
multiplicity density at RHIC. The shadowing factor for quarks and gluons are,
\begin{eqnarray}
R_q^A(x,b)&=&1.0+1.19\log^{1/6}\!\!\!A\;(x^3-1.2x^2+0.21x) \nonumber \\
        &-&s_q(b)\;(A^{1/3}-1)^{0.6}(1-3.5\sqrt{x})\nonumber \\
        &\times& \exp(-x^2/0.01),
\label{eq:shq} \\
R_g^A(x,b)&=&1.0+1.19\log^{1/6}\!\!\!A\;(x^3-1.2x^2+0.21x) \nonumber \\
        &-&s_g(b)\;(A^{1/3}-1)^{0.6}(1-1.5x^{0.35})\nonumber \\
        &\times&\exp(-x^2/0.004),
\label{eq:shg}
\end{eqnarray}
respectively. The impact-parameter dependence of the shadowing is implemented through the parameters,
\begin{equation}
s_a(b)=s_a\frac{5}{3}(1-b^2/R_A^2),
\end{equation}
where $R_A=1.12 A^{1/3}$ is the nuclear size. The value $s_{q}=0.1$ is fixed by the experimental data on
DIS off nuclear targets \cite{Li:2001xa} and $s_{g}=0.17-0.22$ is chosen to fit the RHIC experimental using HIJING 2.0 model
as shown in Fig.~\ref{fig:dndeta-aa}. The form of the impact parameter dependence is chosen to give rise to the
centrality dependence of the pesudorapidity multiplicity density per participant pair  and the range of values of
$s_{g}$ is allowed by experimental errors at the RHIC energies. With this parameterization of parton shadowing,
one can calculate the pesudorapidity multiplicity density per participant pair in $Pb+Pb$ collisions at the LHC
energies  as a function of the number of participants shown in Fig.~\ref{fig:dndeta-aa}. As a comparison we
also show results from HIJING1.0 as dashed lines in Fig.~\ref{fig:dndeta-aa}. The range of the
gluon shadowing parameter $s_{g}$ allowed at RHIC leads to much bigger uncertainties in the multiplicity
density at the LHC energies. With such strong gluon shadowing
constrained by the RHIC data, the effective mini-jet cross section at the LHC energies is still much larger
than at RHIC and therefore the minijet component leads to stronger centrality ($N_{part}$) dependence
of the central pesudorapidity density of produced hadrons. This strong centrality dependence is very different from other model
predictions \cite{Abreu:2007kv} such as the color glass condensate (CGC) or the limiting fragmentation model. Therefore,
the first experimental data on the hadron production at LHC will shed light on the parton/hadron production
mechanism in high-energy hadron and nuclear collisions.

\begin{figure}
  \centering
 \includegraphics[width=0.45\textwidth]{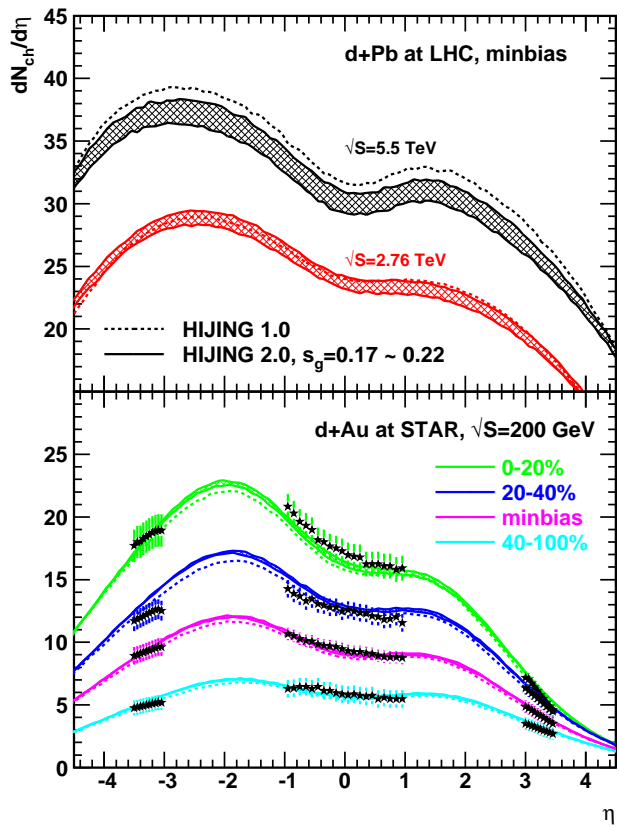}
  \caption{(color online) Pseudorapidity distributions of charged hadrons in $d+Au$ collisions at the
  RHIC energy $\sqrt{s}=200$ GeV/n (lower panel) and $d+Pb$ collisions (upper panel) at the LHC energies.
  The solids lines are HIJING2.0 results with the bands corresponding to the variation of gluon nuclear shadowing
  parameter $s_{g}=0.19-0.24$. Data at the RHIC energy are from STAR experiment \cite{Abelev:2007nt}. }
  \label{fig:dndeta-da}
\end{figure}

To constrain parton shadowing in nucleus, one in principle can also study hadron production in $d+A$ collisions.
Shown in the lower panel of Fig.~\ref{fig:dndeta-da} are the pseudorapidity distributions of charged hadrons
in $d+Au$ collisions at the RHIC energy $\sqrt=200$ GeV for different centralities and minimum-bias events as
compared to the STAR data \cite{Abelev:2007nt}. Within the experimental errors, both HIJING2.0 (solid bands) and HIJING1.0
can describe the data reasonably well. We have followed the definition for the selection of centrality in the STAR
measurements. The solid bands in the HIJING2.0 calculation as the variation of the gluon
shadowing parameter $s_{g}$ constrained by RHIC data of $Au+Au$ collisions are too small to be discerned by
the $d+Au$ data. Similar situation occurs in $d+Pb$ collisions at the LHC energy as shown in the upper panel
of  Fig.~\ref{fig:dndeta-da}. The change of the pesudorapidity distribution due to variation of the gluon nuclear
shadowing parameter is much smaller than in central $Pb+Pb$ collisions shown in Fig.~\ref{fig:dndeta-aa}. The
difference between HIJING2.0 and HIJING1.0 in $d+Pb$ collisions becomes significant only at the highest energy
$\sqrt{s}=5.5$ TeV, reflecting the $p+p$ results in Fig.~\ref{fig:dndet-s}.

\subsection{Conclusions}

We have updated HIJING Monte Carlo model with modern parton distribution functions for the nucleons
and the new set of parameters within the two-component model for mini-jet production in high-energy
nucleon-nucleon collisions. Because of the large gluon distribution at small $x$ in the GRV \cite{Gluck:1994uf}
parameterization of the nucleon's PDFs used in HIJING 2.0, one has to introduce an energy-dependent transverse momentum
cut-off $p_{0}$ for the mini-jet production and the soft parton interaction cross section $\sigma_{soft}$ in order
to describe the energy-dependence of the total, inelastic cross sections and the central rapidity hadron
density in high-energy $p+p (\bar p)$ collisions. The updated HIJING 2.0 model is shown to describe the
existing experimental data on hadron production from ISR energy up to Fermilab Tevatron energy. The
HIJING 2.0 results are also shown to be in good agreement with the recently published hadron spectra in
$p+p$ collisions at the LHC energies ($\sqrt{s}$=0.9, 2.36 TeV), except for events with INEL>0 trigger because
of the lack of double-diffractive events in the model of soft interaction in HIJING. We also give the HIJING 2.0 predictions
for $p+p$ collisions at $\sqrt{s}=7$ and 14 TeV. With a model parameterization for nuclear modification of the parton distribution
functions, we also give HIJING 2.0 prediction of hadron multiplicity in central $Pb+Pb$ and minimum-bias
events of $d+Pb$ collisions  at the LHC energies $\sqrt{s}=2.75$ and 5.5 TeV/n.

This is the first step of the upgrade to the HIJING model. Jet quenching description in current HIJING model
 is very schematic. The next stage of upgrade of HIJING will be focused on jet quenching in dense medium
 incorporating the most recent
 development \cite{Gyulassy:1993hr, Baier:1996sk,Wiedemann:2000za,Gyulassy:2000er,Guo:2000nz}
 in the theory of parton propagation and multiple interaction in dense medium.

\section*{Acknowledgement}

We thank H. Z. Zhang for providing the NLO pQCD results of transverse momentum spectra.
We would like to thank M. Gyulassy for helpful discussions and P. Jacobs and J. Schukraft for discussions
about ALICE experimental data.
This work was supported in part by the National Natural  Science Foundation of China under the project
No. 10525523, No. 10825523, MOE of China under Project No. IRT0624, and the Director, Office of Energy
Research, Office of High Energy and Nuclear Physics, Division of Nuclear Physics, of the U.S. Department of
Energy under Contract No. DE-AC02-05CH11231. W.-T. Deng was also financially supported by Helmholtz
International Center for FAIR within the framework of the LOEWE program launched by the State of Hesse
during the completion of this work.

\end{document}